\documentclass{WileyMSP-template}
\usepackage{graphicx}
\usepackage{amsmath}
\usepackage{amssymb}
\usepackage{siunitx}
\usepackage{hyperref}
\usepackage{physics}
\usepackage{xcolor}

\begin{document}

\pagestyle{fancy}
\rhead{\includegraphics[width=2.5cm]{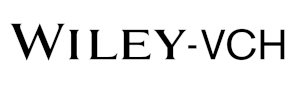}}

\title{High-Fidelity Integrated Quantum Photonic Logic Via Robust Directional Couplers}

\maketitle



\author{Jonatan Piasetzky*$^\dagger$}
\author{Khen Cohen$^\dagger$}
\author{Amit Rotem}
\author{Yuval Warshavsky}
\author{Yehonatan Drori}
\author{Yaron Oz}
\author{Haim Suchowski}

\dedication{}
$\dagger$ Equal contribution

\begin{affiliations}
J. Piasetzky, K. Cohen, Y. Warshavsky, Y. Oz, H. Suchowski\\
School of Physics and Astronomy, Tel Aviv University, 55 Chaim Levanon St., 6997801, Tel Aviv, Israel \\
Email: piasetzky1@mail.tau.ac.il\\
A. Rotem, Y. Drori, Y. Oz, H. Suchowski\\
Quantum Pulse Ventures Ltd. 
\end{affiliations}



\begin{abstract}
  Scalable quantum information processing with integrated photonics requires quantum logic operations with high fidelity and robustness. Directional couplers, the fundamental elements enabling quantum interference and logic operations, are inherently sensitive to fabrication imperfections and environmental fluctuations, leading to reduced gate fidelities. Here, we experimentally demonstrate a passive design strategy that mitigates these errors by exploiting a stationary geometrical configuration in uniform directional couplers, where first-order variations in the coupling coefficient are intrinsically suppressed. The robust geometry is implemented in a silicon-on-insulator photonic chip hosting two-photon controlled-NOT (CNOT) quantum gates and its performance is directly compared to a non-optimized design. Measurements indicate a mean gate fidelity of $93.30 \pm 0.11\%$, representing a clear improvement over the non-robust implementation mean fidelity of $91.93 \pm 0.17\%$, without any active tuning or footprint increase. This performance approaches the theoretical limit of $93.78\%$, imposed by the imperfect source. Monte Carlo simulations incorporating realistic fabrication noise confirm the observed enhancement and reveal consistent suppression of gate-level error rates. These results demonstrate a compact, fabrication-tolerant building block for scalable, fault-tolerant photonic quantum circuits and highlight the power of passive geometric error mitigation in quantum hardware design.
\end{abstract}


\section{Introduction}
Photonic integrated circuits (PICs) have emerged as a promising platform for quantum information processing, offering low-loss manipulation of quantum states, room-temperature operation, and scala
lity through mature semiconductor fabrication technologies~\cite{vlasov_silicon_2012,fang_ultralow_2011,sun_single-chip_2015,xie_heterogeneous_2019,fu_optical_2024}. The implementation of multi-qubit logic on such platforms, however, remains challenging due to the extreme sensitivity of quantum interference to device imperfections. 
Directional couplers, which serve as fundamental building blocks in nearly all quantum photonic circuits and enable both quantum interference and logic operations ~\cite{politi_silica--silicon_2008,silverstone_-chip_2014,matthews_manipulation_2009,fu_quantum_2003}, are particularly susceptible to fabrication and environmental variations. Small deviations in waveguide width, etch depth, waveguide height, or refractive index can shift the coupling coefficient, leading to imbalanced interference, reduced visibility, and increased logical error rates. These effects collectively impede progress toward large-scale, fault-tolerant quantum photonic computing.

Several approaches have been developed to mitigate such errors. Active tuning using thermo-optic or electro-optic phase shifters ~\cite{wang_integrated_2020,rudolph_why_2017,qiang_large-scale_2018} can compensate for fabrication variability, but at the cost of power consumption, thermal drift, and increased circuit complexity. Other strategies, such as adiabatic or asymmetric coupler designs ~\cite{chrostowski_silicon_2015,dong_observation_2006}, expand the fabrication-tolerance window but typically require larger device footprints. More recently, composite and segmented couplers ~\cite{Kyoseva2019,katzman_robust_2022,kaplan_segmented_2023,cohen_robust_2025} have been proposed to realize high-fidelity single-qubit rotations through coherent error cancellation, though these architectures demand precise segmentation and complex phase engineering.

A recently proposed alternative involves stationary-geometry couplers ~\cite{mikkelsen_dimensional_2014}, in which the coupling coefficient becomes locally insensitive to first-order perturbations in fabrication and environmental parameters. This stationary point suppresses variations in the power-coupling ratio arising from changes in waveguide width, providing a simple, footprint-neutral, and fabrication-compatible solution to variability-induced errors. While this effect has been validated in classical optical experiments, its impact at the quantum level, particularly for multi-photon entangling operations, has not yet been experimentally established.

To assess the quantum advantage conferred by stationary-geometry couplers, we evaluate their performance within a two-photon Controlled-NOT (CNOT) gate. The CNOT gate serves as a canonical benchmark for photonic quantum logic ~\cite{raussendorf_topological_2007,wang_confinement-higgs_2003,nielsen_quantum_2010,yariv_photonics_2007}, directly linking device-level precision to logical-level fidelity and enabling a clear comparison between robust and non-robust architectures.

Here, we provide the first experimental demonstration that stationary-geometry couplers improve the fidelity of a multi-photon entangling quantum logic gate. By comparing a robust CNOT circuit to a non-stationary reference circuit fabricated on the same wafer, we isolate the improvements arising solely from passive geometric error suppression. Our measurements show a mean gate fidelity of  93.30±0.11\%, exceeding that of the non-optimized circuit by 1.37\%. This enhancement is consistent with detailed Monte Carlo simulations incorporating experimentally measured coupling-ratio variations.  The improvement reflects a measurable reduction in physical error probability, achieved without active tuning, feedback, or added design complexity. These results validate passive geometric robustness as a compelling route toward scalable and fabrication-tolerant quantum photonic circuits.

\section{Stationary-Geometry Coupler Principle}
\begin{figure}[h]
	\centering
	\includegraphics[width=\textwidth]{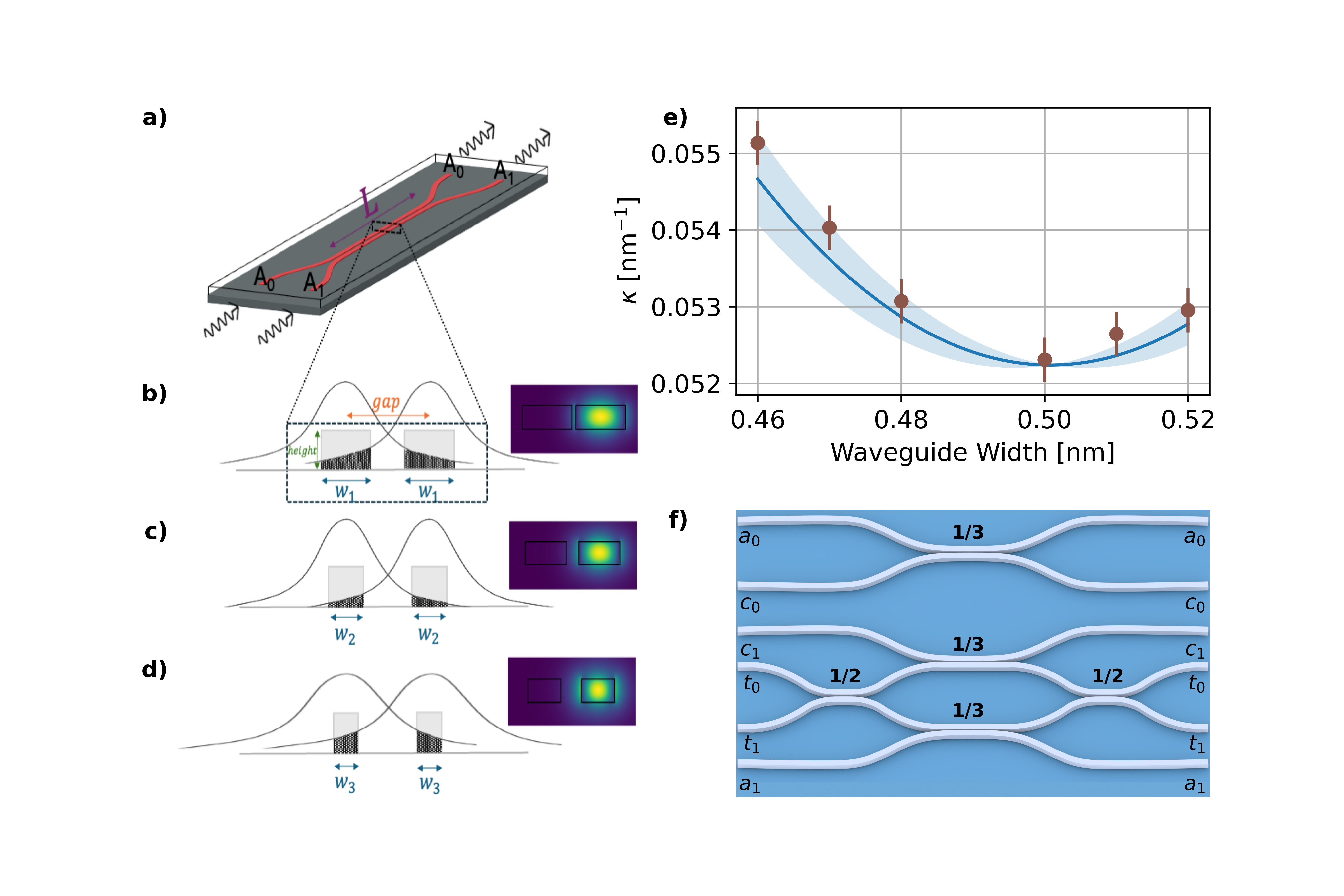}
	\caption{Schematic of the stationary-geometry coupler principle. (\textbf{a}) A schematic drawing of a directional coupler. (\textbf{b-d}) Two-dimensional cross section of the interaction region in a directional coupler, showing three cases. (\textbf{b}) The waveguide widths are wider then the stationary point, resulting in larger coupling coefficient due to large overlap area. (\textbf{c}) Waveguide widths are exactly at the stationary point. (\textbf{d}) The waveguides are narrower than the stationary width, causing a larger coupling coefficient due to stronger evanescent field. (\textbf{e}) The coupling coefficient $\kappa$ is minimized at the stationary width $W_\mathrm{stat}$ as seen from both measurements (brown data points) and numerical simulations (blue solid line). The blue shaded area shows the coupling coefficient variation in the simulation due to a \SI{5}{nm} width variation.(\textbf{f}) The dual-rail CNOT circuit used in this work.} 
	\label{fig:stationary_geometry}
\end{figure}

\begin{figure}[!h]
	\centering
	\includegraphics[width=\textwidth]{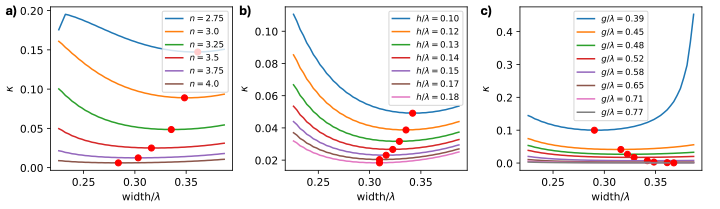}
	\caption{The effect of the refractive index and geometry on the stationary point. The existence of such a stationary point is universal across integrated platforms and wavelengths, though its location depends on the (\textbf{a}) refractive-index contrast and (\textbf{b-c}) specific geometry.}
	\label{fig:critical_width_vs_widths_equalW_wlunits}
\end{figure}

The transfer matrix of a uniform directional coupler can be described by coupled-mode theory (CMT) \cite{yariv_photonics_2007,lifante_integrated_2003}, which captures the coherent exchange of energy between two parallel waveguides. For a symmetric uniform coupler, where both waveguides share identical propagation constants ($\beta_1 = \beta_2$), the transfer matrix for a coupling region of length $L$ is given by $U(L) = \exp(i \kappa L \sigma_x)$, where $\sigma_x$ is the Pauli-X operator and $\kappa$ is the coupling coefficient. According to CMT, $\kappa$ is determined by the overlap of the evanescent fields of the two guided modes. Thus, it depends on several geometrical and material parameters, primarily the waveguide width $W$, inter-waveguide gap $g$, core height $h$, and refractive index $n$. In standard silicon-on-insulator (SOI) platforms, variations as small as 5-\SI{10}{nm} in these quantities, caused by fabrication tolerances, lead to several percent shifts in $\kappa$, directly translating into measurable changes in splitting ratio and, consequently, quantum-gate fidelity. It has been shown \cite{mikkelsen_dimensional_2014} that under a typical fabrication noise scenario, in which the waveguide widths errors are mostly correlated between the two waveguides, while keeping the waveguide centers fixed ($\delta W \approx - \delta g$), this sensitivity can be reduced by engineering the coupler to a stationary point, in which the first order error in $\kappa$, with respect to correlated width and gap variations, vanishes:
\begin{equation}
	\delta \kappa \approx (\frac{\partial \kappa}{\partial W}-\frac{\partial \kappa}{\partial g})\Big|_{W=W_\mathrm{stat}}\delta W \approx 0
\end{equation}
This stationary point emerges from two competing physical effects, mode confinement and proximity effect. On the one hand, increased waveguide width enhances mode confinement, reducing the evanescent overlap between neighboring waveguides and thus decreasing the coupling coefficient. On the other hand, a larger width also decreases the inter-waveguide gap, increasing the overlap region and thereby increasing the coupling coefficient. 
The stationary point itself is manifested at the width in which these two effects cancel out, as shown schematically in Figure \ref{fig:stationary_geometry}\textbf{(a-d)}, producing a locally flat response of the coupling coefficient to fabrication errors. The existence of such a stationary point is universal across integrated platforms and wavelengths, though its location depends on the specific geometry and refractive-index contrast, as seen in Figure \ref{fig:critical_width_vs_widths_equalW_wlunits}. Hence, identifying operating points where the partial derivatives vanish is an effective route to passive error suppression. 

Numerical simulations were carried out using Lumerical MODE, a finite-element eigenmode solver, and were followed by experimental realization on an SOI platform. Both the simulations and the experiments confirm the existence of such a stationary point, at $W_\mathrm{stat}\approx \SI{501}{nm}$, where correlated geometry deviations have minimal effect on the coupling coefficient, as shown in Figure \ref{fig:stationary_geometry}\textbf{(e)}. The cross-section of the interaction region of both the simulated and fabricated devices consisted of two parallel ridge waveguides with a \SI{220}{nm} Si device layer and \SI{2}{\micro\meter} of SiO$_2$ cladding. The gap between the waveguides' centers was fixed at \SI{0.65}{\micro\meter}, and the waveguide widths were scanned to identify the stationary point. 

The simulations were performed using a 2D geometry, and the coupling coefficient was computed from the overlap integral, in the following manner:
$$\kappa = \frac{c}{4\pi\lambda}
\!\iint\!
\big[\epsilon(x,y)-\epsilon_1(x,y)\big]
.
\,E^{(0)}_\perp(x,y)E^{(1)}_\perp(x,y)\,dx\,dy,$$
where $\epsilon_1$ represents the dielectric field with the presence of a single isolated waveguide, and $E^{(1)}_\perp(x,y)$ represents the perpendicular component of the electric field of the single isolated waveguide mode. Similarly, $\epsilon(x,y)$ is the dielectric field of the two-waveguide system, and $E^{(0)}_\perp(x,y)$ represents the perpendicular component of the electric field of the two-waveguide system mode.
The dependence $\kappa(W)$ was fitted with a cubic spline to identify the stationary point $\partial\kappa/\partial W=0$, as shown in Figure \ref{fig:stationary_geometry}\textbf{e} as a blue solid line.

The devices were fabricated through Applied Nanotools multi-project-wafer service. Electron-beam lithography followed by reactive ion etching defined the waveguides geometry. Lastly, a \SI{2}{\micro\meter} of SiO$_2$ cladding was deposited on top. Broadband light from an Erbium-Doped Fiber Amplifier (ASE mode) was injected via a 16-fiber array by PHIX (SMF-28, 127 µm pitch, 40° cleave). Output spectra were recorded using a Yokogawa AQ6370D optical spectrum analyzer. For each coupler, transmission spectra were measured for all four input–output permutations to remove insertion-loss asymmetry \cite{piasetzky_robust_2024}. To experimentally determine the coupling coefficient, multiple devices were fabricated and measured for the same width with varying interaction length. The coupling coefficient was extracted by fitting the power oscillations to the sine function $P_{out}(L)=\sin^2(\kappa L+\theta_c)P_{in}$, where $\theta_c$ accounts for residual coupling that arises from the curved region where the waveguides are brought together. The results are shown in Figure \ref{fig:stationary_geometry}\textbf{e} as brown data points.

\section{Robust Quantum Logic Gates}
Building on this stationary operating point, we directly integrate a fabrication-robust directional coupler into a multi-photon quantum logic circuit. Specifically, we implement the stationary-geometry solution at the level of a single uniform directional coupler and deploy it within a fully integrated dual-rail CNOT gate.

We implemented a fully integrated linear-optical CNOT gate utilizing the post-selected dual-rail architecture proposed in Ref.~\cite{ralph_linear_2002} and previously demonstrated in both bulk~\cite{obrien_demonstration_2003} and integrated systems~\cite{politi_silica--silicon_2008}. As illustrated in Figure \ref{fig:stationary_geometry}\textbf{f}, the device operates via measurement-induced collapse and nonclassical interference, employing directional couplers to execute beam-splitting and merging. The circuit utilizes dual-rail path encoding, where each qubit consists of two spatial modes representing the logical $\ket{0}$ and $\ket{1}$ states. Consequently, the two-qubit system comprises four waveguides labeled $c_{0,1}$, and $t_{0,1}$, which denote the control and target qubits, respectively, and two additional ancilla waveguides labeled $a_{0,1}$. The network design incorporates five directional couplers - two with reflectivity $R=1/2$ and three with $R=1/3$. These couplers apply unitary transformations defined by $U_R = \exp(i \sigma_x \phi_R)$, where $\sigma_x$ denotes the Pauli-X operator and $\phi_R=\arccos{\left(\sqrt{R}\right)}$ is the coupling angle of a coupler with reflectivity $R$.

Post-selection on coincident detection yields the effective unitary transformation up to a global phase:
$$|00\rangle \!\to\! |00\rangle,\qquad
|01\rangle \!\to\! |01\rangle,\qquad
|10\rangle \!\to\! |11\rangle,\qquad
|11\rangle \!\to\! |10\rangle$$

To experimentally verify the quantum-level benefit, we fabricated two devices on the same multi-project wafer run, ensuring identical process conditions. The first device is a CNOT gate with all couplers engineered at $W_\mathrm{stat}= \SI{501}{nm}$, while the second device is a CNOT gate with all couplers fabricated at an arbitrary width of $\SI{450}{nm}$. In all couplers, the interaction length was tuned to achieve the required reflectivity.

The quantum characterization used photon pairs generated by spontaneous parametric down-conversion (SPDC) in a periodically poled KTP crystal pumped by a continuous-wave diode laser at wavelength of \SI{775}{nm}.
The degenerate signal and idler photons at \SI{1550}{nm} were filtered with a \SI{12}{nm} Bragg filter and coupled into polarization-maintaining fibers before entering the on-chip input grating couplers.
Coincidence detection was performed using superconducting nanowire single-photon detectors (SNSPDs) connected to a time-correlated single-photon counting (TCSPC) module.

\section{Results}
\begin{figure}[h]
	\centering
	\includegraphics[width=\textwidth]{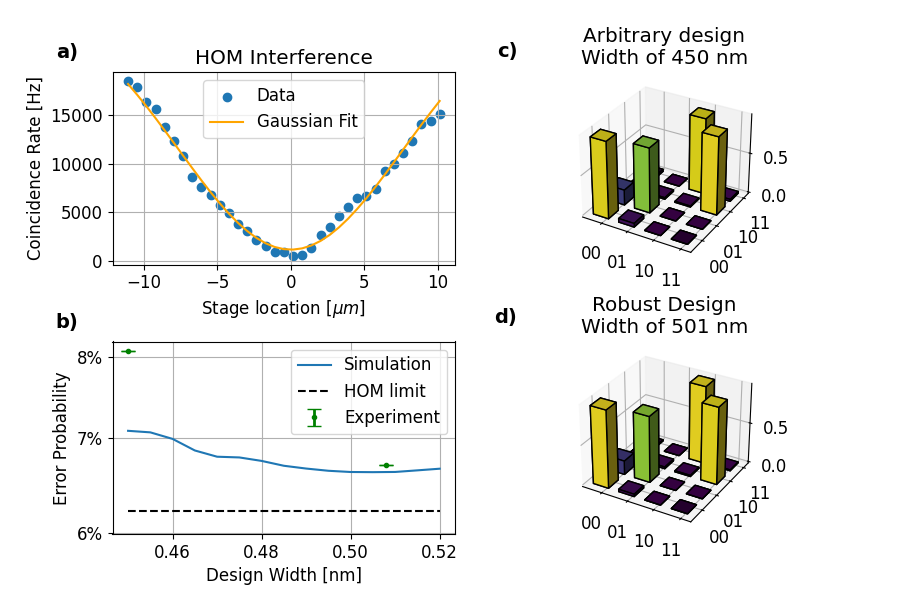}
	\caption{Truth table reconstruction of the CNOT gate. \textbf{a} The Hong-Ou-Mandel interference profile of the CNOT gate. The visibility was extracted by fitting a Gaussian function (orange) to the HOM interference profile (blue), and was reported as 95.7\%. \textbf{b} The mean error probability as a function of the design width. The blue curve shows the error probability as a function of the design width as obtained from Monte Carlo simulations. The experimental results are shown as green points, and show even better improvement than the theoretical predictions. The HOM limit was calculated to be 6.22\% and is shown as a black dashed line. \textbf{c} The truth table reconstruction of the CNOT gate for the reference device. It shows a mean error probability of 8.07\% ± 0.17\%. \textbf{d} The truth table reconstruction of the CNOT gate for the robust design. It shows a mean error probability of 6.70\% ± 0.11\%.}
	\label{fig:cnot_truth_table}
\end{figure}

The performance of linear-optical quantum gates critically depends on high-visibility interference between indistinguishable photons.
Prior to circuit testing, we verified photon indistinguishability using a separate Hong–Ou–Mandel (HOM) interferometer. The measurement results are shown in Figure \ref{fig:cnot_truth_table}\textbf{a}. The visibility was extracted by fitting a Gaussian function to the HOM interference profile. The measured visibility of $95.7\%$ indicates photon distinguishability arising from source imperfections. It also limits the maximal fidelity of the CNOT gate to $ 93.78\% $ ~\cite{nakav_quantum_2025-1}. For the simulations, this distinguishability was also included as a non-interfering mixed state contribution in the numerical model. The HOM interference profile was measured to be Gaussian with a 12 nm full-width-at-half-maximum, consistent with the transmission spectrum of the on-chip grating couplers.

We characterized both circuit variants, the robust design using stationary-geometry couplers and the reference design with arbitrary width, under identical conditions. For each logical input state $|00\rangle, |01\rangle, |10\rangle, |11\rangle$, coincidence counts were accumulated for all four logical output ports. The raw counts were background-subtracted, normalized to total coincidences per input, and corrected for detector-efficiency imbalance. The resulting probability matrices, which are shown in Figure \ref{fig:cnot_truth_table}\textbf{c-d}, reproduce the characteristic behavior of a post-selected CNOT gate: the target qubit flips only when the control qubit is in the logical $|1\rangle$ state.

We evaluated the mean error probability as the probability to measure the incorrect output state, averaged over all input states. The results are shown in Figure \ref{fig:cnot_truth_table}\textbf{b}. The robust design achieved a mean error probability of 6.70\% ± 0.11\%, compared with 8.07\% ± 0.17\% for the reference device. This corresponds to about 17\% reduction in mean error probability, consistent with the expected suppression of first-order coupling-ratio variations at the stationary point. Relative to the remaining error budget set by photon distinguishability, this corresponds to a reduction of almost fourfold in the error probability.

Monte-Carlo simulations incorporating measured $\kappa(W)$ dependencies, spectral effects, and the 95.7\% HOM visibility are shown in Figure \ref{fig:cnot_truth_table}\textbf{b}, and qualitatively reproduce the stationary-geometry effect in the quantum two-qubit CNOT gate.

The demonstrated fidelity improvement is achieved entirely passively, without thermo-optic tuning, feedback, or additional footprint. Given that physical-error thresholds for surface-code architectures lie near the 1\% level, each percentage point of intrinsic fidelity gain significantly reduces the overhead of active error correction. Moreover, the stationary-geometry concept applies universally to any two-waveguide coupler, making it directly transferable to alternative material systems such as SiN or LiNbO$_3$.

\begin{figure}[!h]
	\centering
	\includegraphics[width=\textwidth]{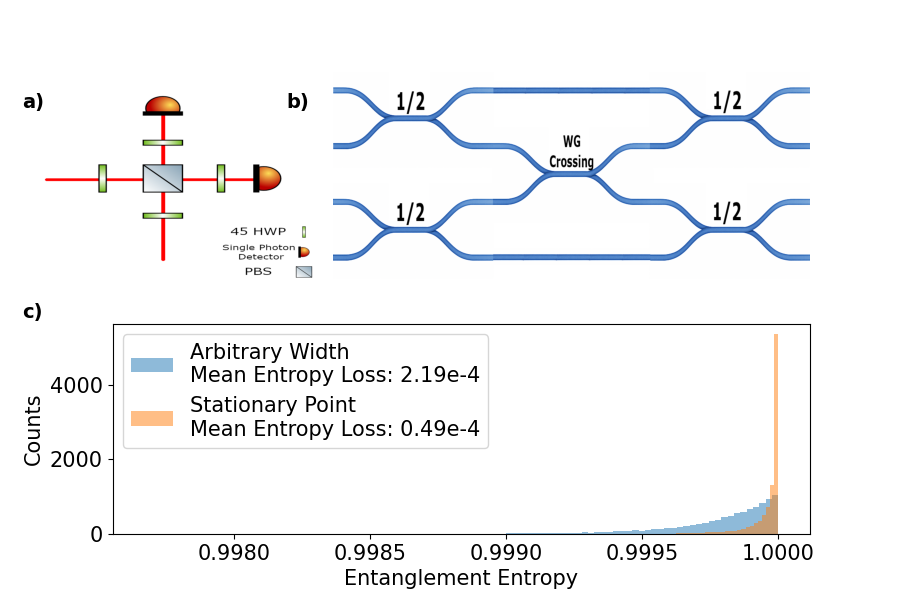}
	\caption{The effect of stationary-geometry couplers on the resource generation for one-way quantum computing. \textbf{(a)} The Type-II fusion gate in its well known free space implementation, which consists of a single polarizing beam splitter (PBS) and four half-wave plates (HWP) rotated at a 45 degree angle. \textbf{(b)} The equivalent integrated dual-rail circuit, which consists of four directional couplers with a reflectivity of 1/2 and a single waveguide crossing. \textbf{(c)} The robust design (orange) achieves a mean entanglement entropy loss of 0.49 \texttimes $10^{-4}$, as opposed to the entropy loss of the arbitrary design (blue) of 2.19 \texttimes $10^{-4}$. This demonstrates the possible usage of stationary-geometry couplers in highly entangled state generation.}
	\label{fig:fusion_resource_generation}
\end{figure}

Furthermore, we examine the effect of the stationary-geometry on the resource generation for one-way quantum computing using Monte Carlo simulations. Large cluster states are a fundamental building block for one-way quantum computing ~\cite{bartolucci_fusion-based_2021,kok_five_2009,rudolph_why_2017,raussendorf_one-way_2001,raussendorf_measurement-based_2003}. The generation of such states is one of the main challenges for one-way quantum computing today. One way to produce such states is by utilizing the well-known fusion gates ~\cite{kok_five_2009}. These gates are used to fuse together two smaller cluster states into a single larger entangled cluster state. One such fusion gate is called the Type-II fusion gate and is shown in its well-known free space implementation in Figure \ref{fig:fusion_resource_generation}\textbf{a}. Assuming two ideal Bell pairs, with a single photon from each pair input into the fusion gate, the expected output state is a maximally entangled Bell pair made from the two paired photons, with an entanglement entropy of 1. To evaluate the effect of stationary-geometry couplers on the resource generation for one-way quantum computing, we created an equivalent integrated dual-rail circuit (Figure \ref{fig:fusion_resource_generation}\textbf{(b)}). Next, we performed the simulations by constructing 1000 instances of type-II fusion gates with and without stationary-geometry couplers. For that we used the measured performance of a single coupler, and evaluated its effect on the entanglement generation of an entire fusion operation by measuring the resulting state entanglement entropy. We found that the stationary-geometry couplers can significantly improve the resource generation fidelity, an entanglement entropy loss of \num{0.49e-4}, as opposed to the reference device with an entanglement entropy loss of \num{2.19e-4}, more than four times higher, as shown in Figure \ref{fig:fusion_resource_generation}\textbf{(c)}. This demonstrates the possible usage of stationary-geometry couplers in highly entangled state generation. The results are summarized in Table \ref{tab:performance_data}.

\begin{table}[h!]
	\centering
	\begin{tabular}{|c|c|c|c|}
		\hline
		& Reference Design (450 nm) & Robust Design (501 nm) & Improvement \\ 
		\hline
		Type-II Fusion Mean Entropy Loss & $2.19 \times 10^{-4}$  & $0.49 \times 10^{-4}$  & $>4\times$ reduction  \\
		\hline
		CNOT Mean Error Probability & $8.07 \pm 0.17\%$ & $6.70 \pm 0.11\%$ & $\approx 17\%$ reduction \\
		\hline
		HOM limit & \multicolumn{3}{c|}{6.22\%} \\
		\hline
	\end{tabular}
	\caption{Performance comparison of reference vs. stationary-geometry quantum gates.}
	\label{tab:performance_data}
\end{table}

\section{Discussion and Conclusion}
Our results demonstrate that stationary-geometry directional couplers provide a direct and practical route to reducing gate-level errors in integrated quantum photonic circuits. By nullifying the first-order sensitivity of the coupling coefficient to fabrication-induced geometric variations, the stationary point effectively flattens the coupler transfer function, converting dominant process fluctuations into higher-order perturbations. This passive robustness leads to measurable improvements in quantum-gate fidelity without relying on active tuning, feedback control, footprint expansion, or additional power consumption.

Experimentally, we observe a reduction in the mean error probability of a two-photon CNOT gate from 8.07\% ± 0.17\% to 6.70\% ± 0.11\%, corresponding to a 17\% improvement relative to a non-optimized design fabricated under identical conditions. Monte Carlo simulations incorporating realistic fabrication noise and measured photon indistinguishability confirm that this gain arises from suppressed first-order coupling-ratio variations. Importantly, the observed improvement approaches the fundamental fidelity limit imposed by the photon source, highlighting the effectiveness of geometric error mitigation at the component level.

Compared with existing robustness strategies, including composite or segmented couplers, adiabatic designs, and active phase trimming, the stationary-geometry approach offers a complementary advantage. It relies solely on intrinsic device geometry, making it fully compatible with standard foundry processes and scalable to large, densely integrated circuits. Moreover, stationary-geometry couplers can be directly embedded within more complex passive or active error-mitigation schemes, reducing their baseline sensitivity and overall control overhead.

Beyond two-qubit logic, we show through numerical simulations that stationary-geometry couplers significantly enhance the fidelity of resource generation for one-way quantum computing. In particular, their use in integrated Type-II fusion gates reduces entanglement entropy loss by more than a factor of four, directly improving the efficiency and stability of cluster-state growth, an essential requirement for photonic fault-tolerant architectures.

In summary, we establish a clear link between classical geometric optimization of photonic components and quantum-logic performance. By embedding robustness into the physical layout of directional couplers, we demonstrate a compact, CMOS-compatible, and purely passive strategy for enhancing quantum-gate fidelity. The stationary-geometry principle is independent of wavelength, material platform, and circuit topology, and can be readily extended to other quantum photonic elements such as interferometers, Hadamard gates, and fusion-based entanglement modules for quantum computing applications, as well as QKD and quantum networks. These results provide a scalable pathway toward high-fidelity, resource-efficient, and ultimately fault-tolerant integrated photonic quantum computing.
\medskip

%
\bibliographystyle{MSP}
\bibliography{references.bib}

\end{document}